# Some comments on the reliability of NOAA's Storm Events Database

**Renato P. dos Santos**

ULBRA - Lutheran University of Brazil
PPGECIM - Doctoral Program in Science and Mathematics Education

June, 22, 2016

## Abstract

Storms and other severe weather events can result in fatalities, injuries, and property damage. Therefore, preventing such outcomes to the extent possible is a key concern, and the scientific community faces an increasing demand for regularly updated appraisals of evolving climate conditions and extreme weather. NOAA's *Storm Events Database* is undoubtedly an invaluable resource to the general public, to the professional, and to the researcher. Due to such importance, the primary objective of this study was to explore this database and get clues about its reliability. A complete investigation of the damage estimates, injuries or fatalities figures is unfeasible due to the extension of the database. However, an exploratory data analysis with the resources of the R statistical data analysis language found that damage reports are missing in more than half of the records, that part of the damage values are incorrect, and that, despite all efforts of standardizations, non-standard event type names are still finding their way into the database. These few results are enough to demonstrate that the database suffers from incompleteness and inconsistencies and should not be used without taking reservations and appropriate precautions before advancing any inferences from the data.

## Introduction

Storms and other severe weather events can cause both public health and economic problems for communities and municipalities. Many serious events can result in fatalities, injuries, and property damage, and preventing such outcomes to the extent possible is a key concern.

According to Trenberth, Fasullo, & Shepherd (2015), "the main way in which climate change is likely to affect societies around the world is through changes in extremes. As a result, the scientific community faces an increasing demand for regularly updated appraisals of evolving climate conditions and extreme weather. Such information would be immensely beneficial for adaptation planning."

This study evolved from an assessment during the *Exploratory Data Analysis* course, which is part of Johns Hopkins *Data Science Specialization* provided by Coursera. It involved exploring the version 1.0 of U.S. National Oceanic and Atmospheric Administration's (NOAA) *Storm Events Database*, comprising data from 1950-01-03 to 2011-11-30. This database tracks characteristics of major storms and weather events in the United States, including when and where they occur, as well as estimates of any fatalities, injuries, and property damage. It is fed with information from a variety of sources, which include but are not limited to county, state and federal emergency management officials, local law enforcement officials, Skywarn spotters, NWS damage surveys, newspaper clipping services, the insurance industry and the general public (NCDC, 2008).

*Storm Data* is an official monthly publication of the National Oceanic and Atmospheric Administration (NOAA) which documents (Murphy, 2016, p. 4):

- The occurrence of storms and other significant weather phenomena having sufficient intensity to cause public health and/or economic problems
- Rare, unusual, weather phenomena that generate media attention
- Other significant meteorological events.

This database is updated monthly and generally lags 90-120 days behind the current month, what makes it a very accessible data source. It constitutes an invaluable resource that is heavily used by the general public, insurance adjusters, litigators, and severe weather climatologists. Thousands of scientific papers have been made using this database. For recent examples, see Miller, Black, Williams, & Knox (2016) and Schroeder et al. (2016).

The primary goal of this study was to explore NOAA's *Storm Events Database* and get clues about its reliability.

## Data

The NOAA's *Storm Event Database* is available in the form of comma-separated-value (.csv) files compressed via the *bzip2* algorithm to reduce their size. They can be downloaded from the NOAA web site. The files are named in the form

**StormEvents_details-ftp_v1.0_dYYYY_c20160223.csv**,

where **YYYY** indicates the year of the events records.

Inspection shows that the database has the following 48 variables: BEGIN_YEARMONTH, BEGIN_DAY, BEGIN_TIME, END_YEARMONTH, END_DAY, END_TIME, EPISODE_ID, EVENT_ID, STATE, STATE_FIPS, YEAR, MONTH_NAME, EVENT_TYPE, CZ_TYPE, CZ_FIPS, CZ_NAME, WFO, BEGIN_DATE_TIME, CZ_TIMEZONE, END_DATE_TIME, INJURIES_DIRECT, INJURIES_INDIRECT, DEATHS_DIRECT, DEATHS_INDIRECT, DAMAGE_PROPERTY, DAMAGE_CROPS, SOURCE, MAGNITUDE,

MAGNITUDE_TYPE, FLOOD_CAUSE, CATEGORY, TOR_F_SCALE, TOR_LENGTH, TOR_WIDTH, TOR_OTHER_WFO, TOR_OTHER_CZ_STATE, TOR_OTHER_CZ_FIPS, TOR_OTHER_CZ_NAME, BEGIN_RANGE, BEGIN_AZIMUTH, BEGIN_LOCATION, END_RANGE, END_AZIMUTH, END_LOCATION, BEGIN_LAT, BEGIN_LON, END_LAT, END_LON, EPISODE_NARRATIVE, EVENT_NARRATIVE, and DATA_SOURCE. They are described at (NOAA, 2014).

# Data Preparation

This work was done with the resources of the R statistical data analysis language (R Core Team, 2016) R version 3.2.5 (2016-04-14) on Windows 7 (build 7600). All efforts were made to conform to the best practices of Reproducible Research (Peng, 2011, 2016a).

After installing needed packages *ggplot2* (Wickham, 2009), *gridExtra* (Auguie, 2016), *scales* (Wickham, 2016), *readr* (Wickham & Francois, 2015a), *dplyr* (Wickham & Francois, 2015b), *knitr* (Xie, 2015), *lubridate* (Grolemund & Wickham, 2011), *XML* (Lang & The CRAN Team, 2016), and *pander* (Daróczi & Tsegelskyi, 2015), we begin by downloading the storm database event details .csv files, from the NOAA website. Afterwards, they are read into R.

The only relevant variables for this study are EPISODE_ID, YEAR, MONTH_NAME, EVENT_TYPE, DAMAGE_PROPERTY, DAMAGE_CROPS, EPISODE_NARRATIVE, and, therefore, only these were read into the dataset.

For convenience, we make the variable names more readable as *EpisodeID*, *Year*, *Month*, *EventType*, *PropertyDamage*, *CropDamage*, and *Narrative*.

# Analysis

Before any further processing, it is advisable to check if the dataset needs some data cleaning, which is considered an essential part of the statistical analysis (de Jonge; van der Loo, 2013). We will check if the dataset lacks headers, contains wrong data types (e.g. numbers stored as strings), bad category labels, unknown or unexpected character encoding and so on.

## Event type names

Now, according to (Murphy, 2016), "the chosen [type] event name should be the one that most accurately describes the meteorological event leading to fatalities, injuries, damage, etc. However, significant events, such as tornadoes, having no impact or causing no damage, should also be included in Storm Data."

The database currently contains data from January 1950. However, in the earlier years of the database the coverage of events was reduced. According to the NOAA website, from 1950 through 1954, only tornado events were recorded; from 1955 through 1995, only tornado, thunderstorm wind, and hail events were introduced into the database; from 1996 to present, all event types as defined in *NWS Directive 10-1605* are recorded.

According to the NOAA website, data before 1995 had many inconsistencies in the spelling of event types and most of them were standardized into the 48 current event types at NCEI in 2013. From 1996-1999, the event type field was a free-text field, so there were many, many variations of event types. In 2000 the NWS added a drop-down selector for Event Type on the data entry interface, which standardized the Event Type values. However, the present *NWS Directive 10-1605* (Murphy, 2016) mentions that seven new events were introduced: Marine Dense Fog, Marine Heavy Freezing Spray, Marine Hurricane/Typhoon, Marine Lightning, Marine Tropical Depression, Marine Tropical Storm, and Sneaker Wave. Accordingly, Table 1 of Section 2.1.1 of lists 55 event names, from *Astronomical Low Tide* to *Winter Weather*.

Therefore, the variable *EventType* was expected to have 48 or 55 distinct values. However, inspection shows that there are 73 distinct values in this variable. Nevertheless, this is quite an improvement as the study mentioned above of its version 1.0 found not less than 950 different values including multiple event types in the same record e.g. "Thunderstorm Wind/Hail" or "Hail/Tornado", a fact already observed by NOAA in 2012 and that implied in "manual edits" "made to split records that contained multiple event types in the record" (NOAA, n.d.).

Figure 1 shows that, despite all these standardizations, non-standard event names have been introduced since 1993, with a peak of 10 non-standard names on 1995, followed by an almost total pause until 2006, and then uninterruptedly up to now.

**Figure 1 - Number of non-standard event names introduced**

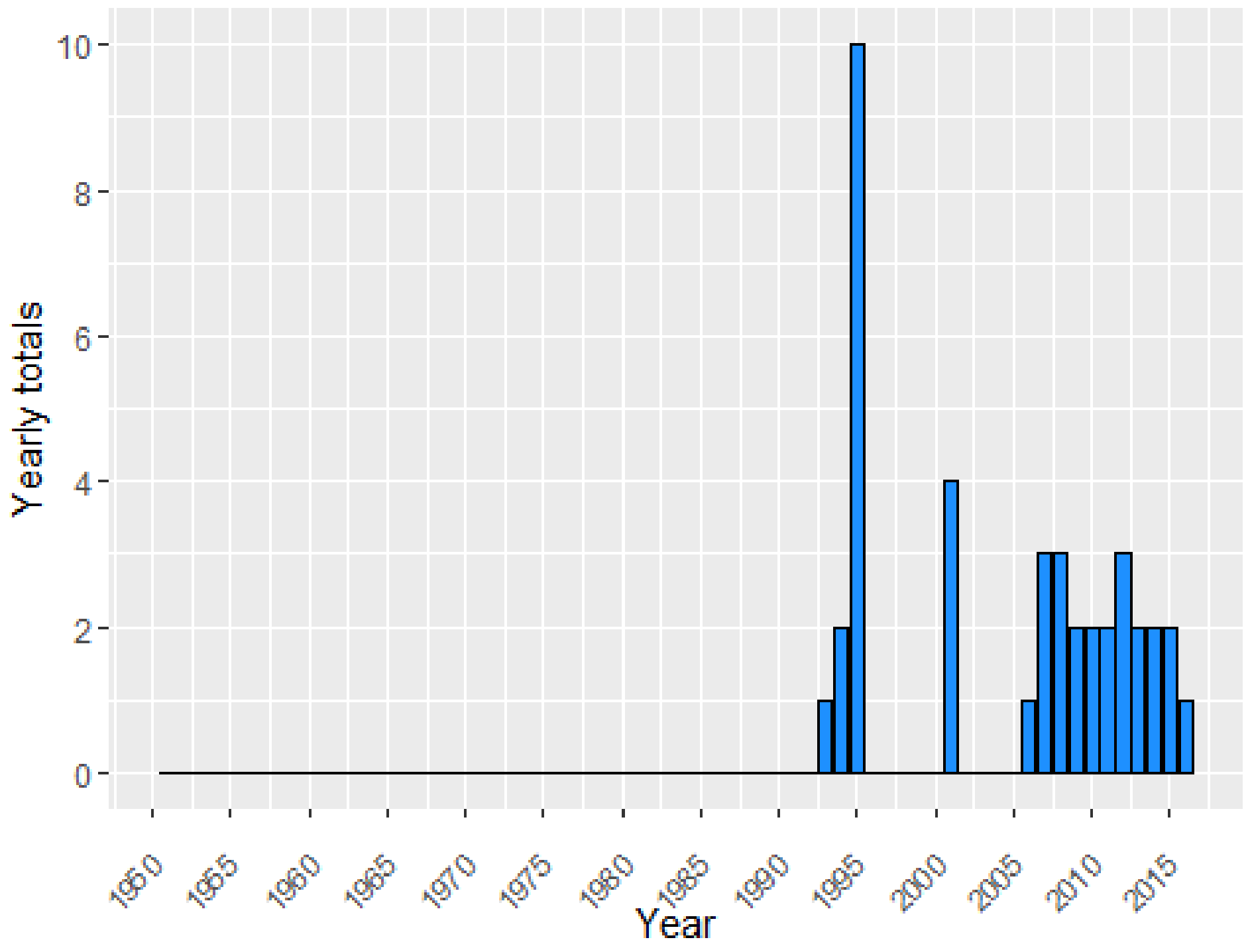

Figure 2 shows the number of total event names (including both standard and non-standard ones) raising up to the present total of 73.

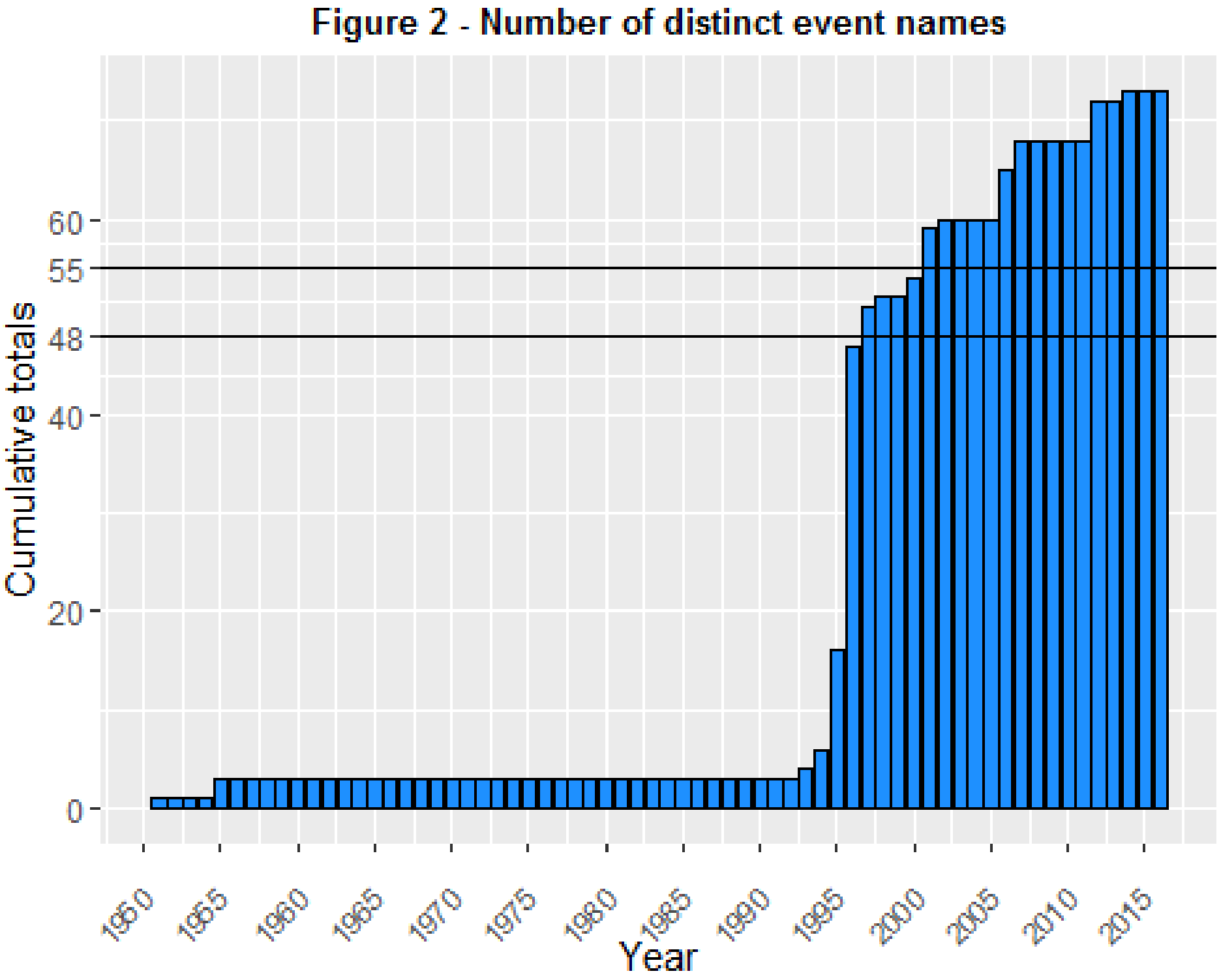

Figure 3 shows that, fortunately, the presence of non-standard event type names in the dataset is quite irrelevant, amounting to no more than one-fifth of a percent of the total number of events each year.

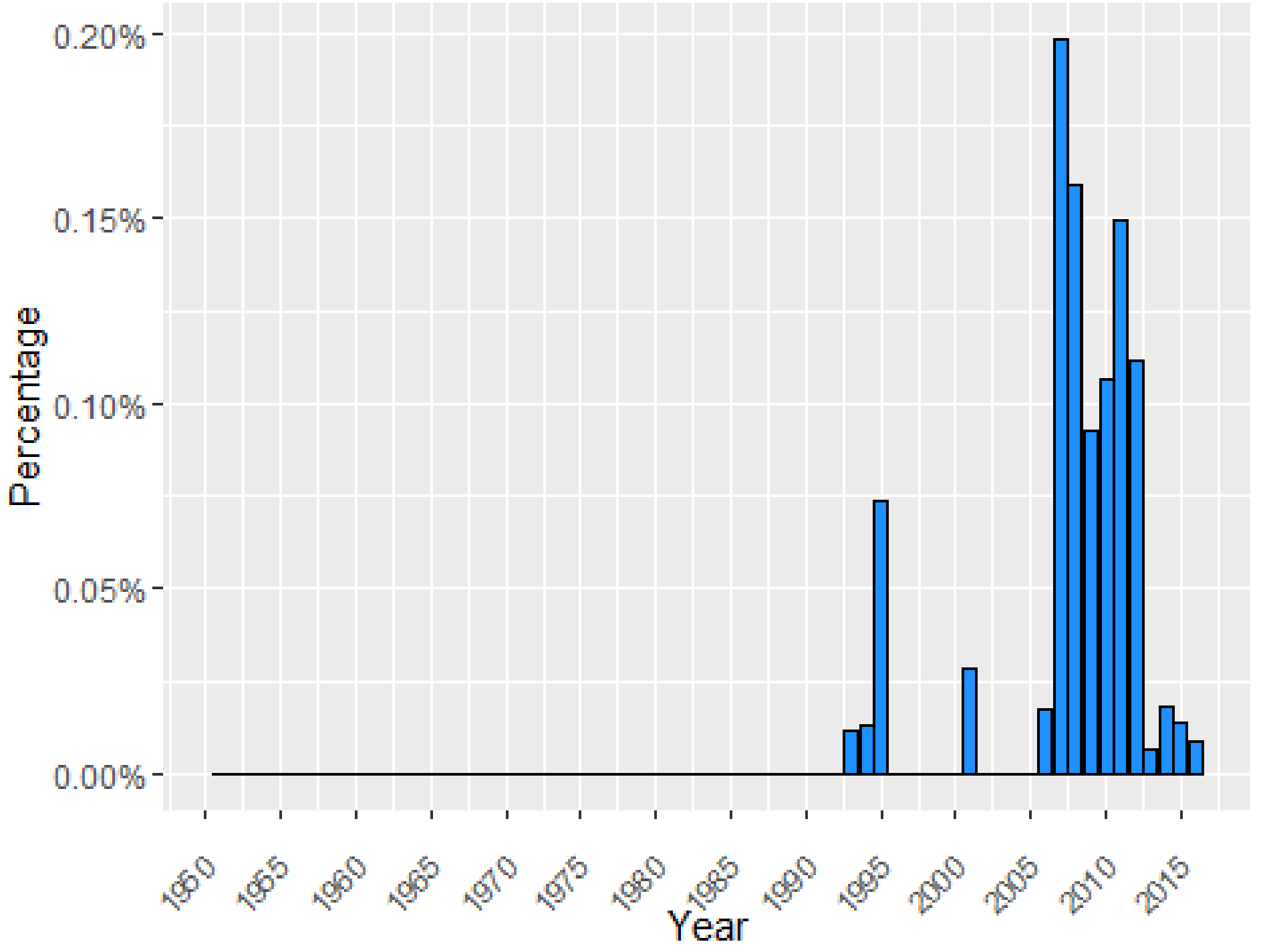

Figure 3 - Percentages of non-standard event type names in the datab

## Damage values

Property damage refers to damage inflicted to private property (structures, objects, vegetation) as well as public infrastructure and facilities.

Tornadoes may contain multiple segments and are reported in Storm Data in separate segments (NCDC, 2008). This fact may affect the attribution of harmful effects for population health and to the economy to individual tornadoes.

According to *NWS Instruction 10-1605* (Murphy, 2016), estimates should be in the form of US Dollar values and rounded to three significant digits, followed by the magnitude of the value (i.e., 1.55B for $1,550,000,000). Values used to signify magnitude include: *K* for thousand $USD, *M* for million $USD, and *B* for billion $USD. Inspection shows, however, that many other values were used as magnitudes: K, M, 0, 3, B, 4, 2, 6, h, 5, 1, H, 9, 7, 8. It is, of course, impossible to be completely sure what the preparer had in mind when introducing these values, but we can make educated guesses about their meaning, such as *h* for hundreds and *6* for millions. Now, we recalculate *PropertyDamage* and *CropDamage* variables taking these magnitude values into account.

Having the damages values in a proper form, it is interesting to check for missing values (coded as *NA*). It is known that missing values are a problem that plagues any data analysis, their presence introducing bias into some calculations or summaries of the data (Peng, 2015, p. 134).

**Figure 4 - Percentages of missing values in the variables.**

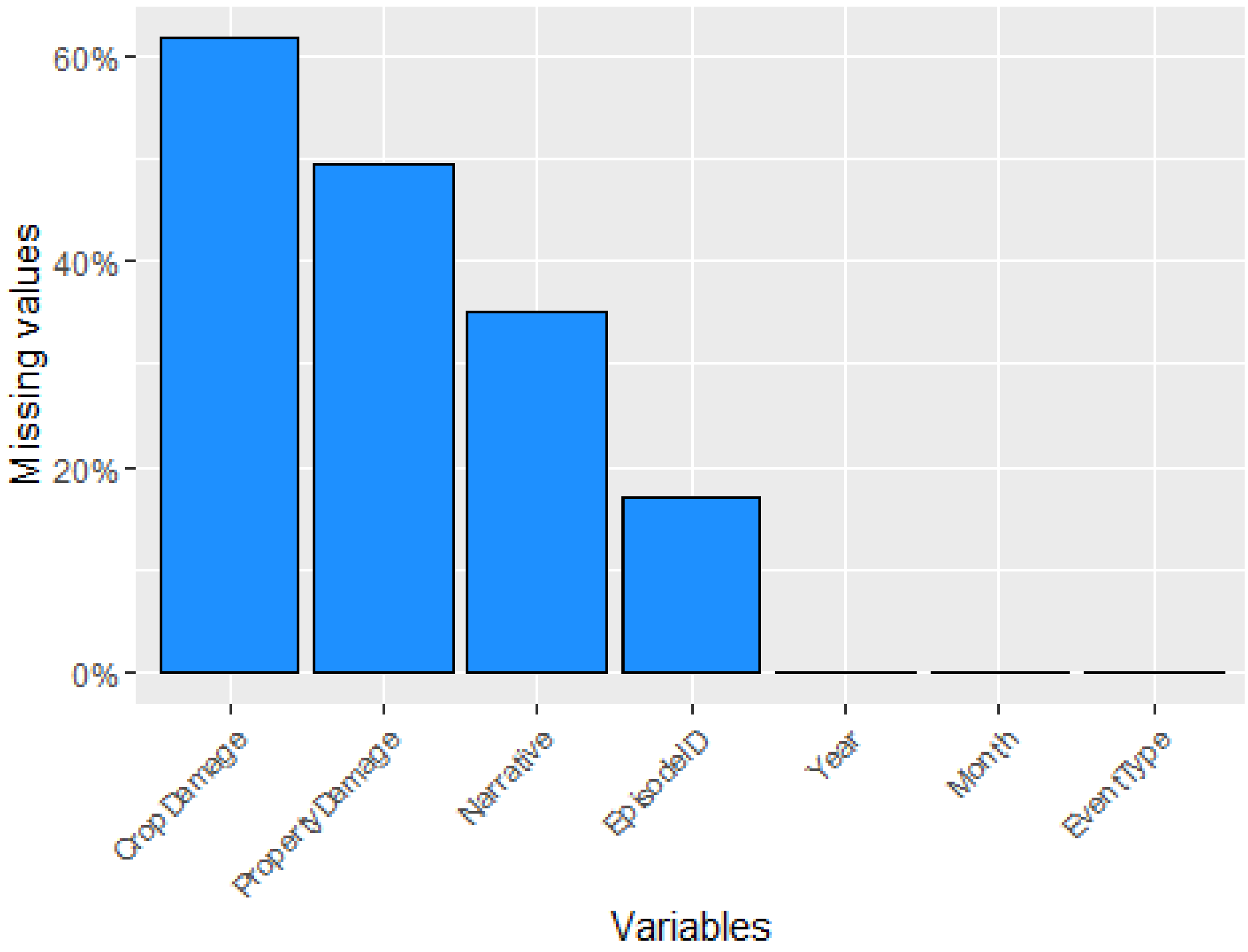

Figure 4 shows that this version of NOAA's *Storm Event Database* is quite faulty in terms of property damages and crop damages values, with the percentage of missing values in the *CropDamage* variable rising up to 62%. This incompleteness is surprising as the study mentioned above of its version 1.0 found no missing values in these variables, but lots of them in the magnitude character code. A reasonable hypothesis is that the complete rebuild of the database that happened in 2012 discarded all damage values without definite magnitudes.

Figure 5 summarizes their values.

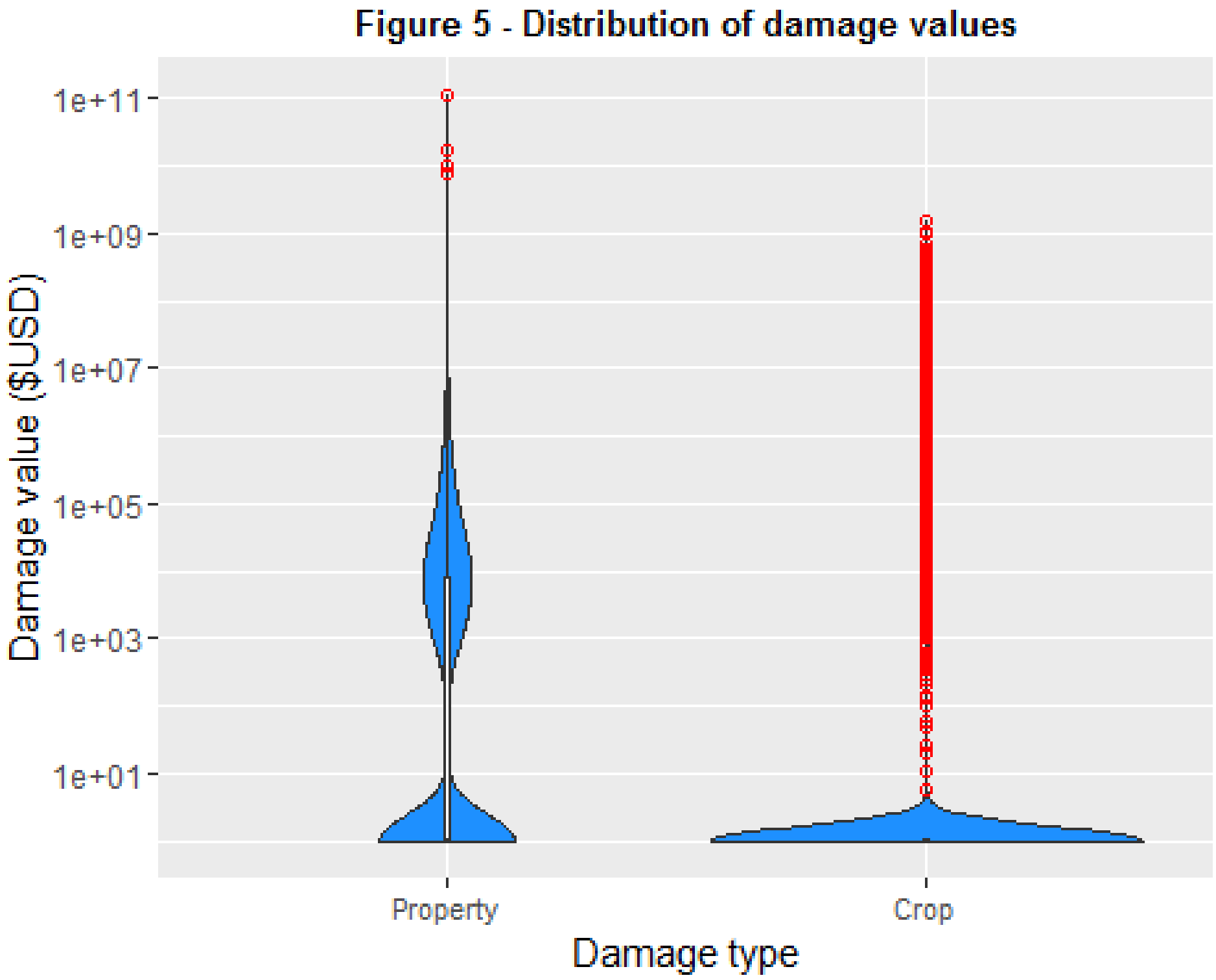

The violin plots (Hintze & Nelson, 1998) of Figure 5 show that property and crop damage values concentrate below *1*, or about US$0.00 if we notice that, due to the logarithmic scale used, the damage values had to be incraesed by US$1.00 (see code in the Appendix).

Figure 5 also exhibit an outlier property damages value of $1.15 \times 10^{11}$, that is, US $115 billion, which is higher than Katrina, estimated at US $108 billion, considered the most destructive and costliest natural disaster in the history of the United States (Knabb, Rhome, Brown, 2011). Inspection shows that this entry corresponds to an event of type *FLOOD* that would have occurred on January 1, 2006, in Napa, CA. The event narrative states: "Major flooding continued into the early hours of January 1st, before the Napa River finally fell below flood stage, and the water receded. Flooding was severe in Downtown Napa from the Napa Creek, and the City and Parks Department was hit with $6 million in damage alone. The City of Napa had 600 homes with moderate damage, 150 damaged businesses with costs of at least $70 million."

Further investigation shows that a flood did take place in Napa on that date, but was "not as bad at the devastating 1986 storm that caused $100 million in damage" (Courtney, 2005). As a matter of fact, this event does not show up in NOAA's *Billion-Dollar Weather and Climate Disasters: Table of Events* webpage.

A possible interpretation is that the preparer has introduced the magnitude character *B* for billions, instead of *M* for millions, what seems more reasonable given the costs "of at least $70 million," mentioned above.

Now, let us introduce a new *TotDamage* variable accounting for the total damage. Table 1 lists the 10 top costliest events recorded in NOAA's *Storm Events Database*.

*The top 10 costliest events (in Billion $USD)*

| EpisodeID | Year | Month | TotDamage |
|---:|---:|:---:|---:|
| 1203478 | 2006 | January | 115 |
| 68471 | 2012 | October | 25 |
| 1198432 | 2005 | August | 7 |
| 1181034 | 2004 | September | 5 |
| 50455 | 2011 | April | 3 |
| 49972 | 2011 | May | 3 |
| 64742 | 2012 | July | 2 |
| 53127 | 2011 | May | 2 |
| 2084790 | 1998 | September | 2 |
| 89491 | 2014 | August | 2 |

Inspection of the corresponding episode *narrative* field, followed by searches on NOAA's *Billion-Dollar Weather and Climate Disasters: Table of Events* webpage filtered from 2001 on help identify these events and verify the provided damage amounts.

*Tentative identification of the top 10 costliest events*

| EpisodeID | Year | Month | Event damage | Name | Listed damage |
|---:|---:|:---:|---:|:---|---:|
| 1203478 | 2006 | January | 115 | Napa River Flood 2005/2006 | 0.1 |
| 68471 | 2012 | October | 25 | Post Tropical Storm Sandy | 65 |
| 1198432 | 2005 | August | 7 | Hurricane Katrina | 125 |
| 1181034 | 2004 | September | 5 | Hurricane Frances | 10 |
| 50455 | 2011 | April | 3 | Southeast/ Ohio Valley/ Midwest Tornadoes | 10 |
| 49972 | 2011 | May | 3 | Midwest/ Southeast Tornadoes | 9 |
| 64742 | 2012 | July | 2 | U.S. Drought/Heatwave | 30 |
| 53127 | 2011 | May | 2 | Mississippi River flooding | 3 |
| 2084790 | 1998 | September | 2 | Hurricane Georges | 6 |
| 89491 | 2014 | August | 2 | Michigan and Northeast Flooding | 1 |

It should be noticed, however, that the amounts shown in Table 2 refer to *episodes* damages, which may be part of a bigger, more extense storm system, such as the Hurricane Katrina, that can include many different types of events, and, therefore, differ from the total damage estimates shown in the last column of this table.

Besides, a few Billion-Dollar events are missing in comparison with the *Billion-Dollar Weather and Climate Disasters: Table of Events* webpage, such as: the 2012 U.S. Drought/Heatwave ($33.3 billion), the 2004 Hurricane Ivan ($25.8 billion), the 2005 Hurricane Wilma ($23.2 billion), and the 2005 Hurricane Rita ($22.6 billion).

On the other hand, except the Napa River Flood 2005/2006, the other episode estimates are reasonably lower than the corresponding total estimates.

This observation is comforting but does not ensure by itself that the remaining estimate values in NOAA's *Storm Events Database* are all trustable.

Figure 6 lists the 5 most frequent event types in the database.

**Figure 6 - The 5 most frequent event types in the database.**

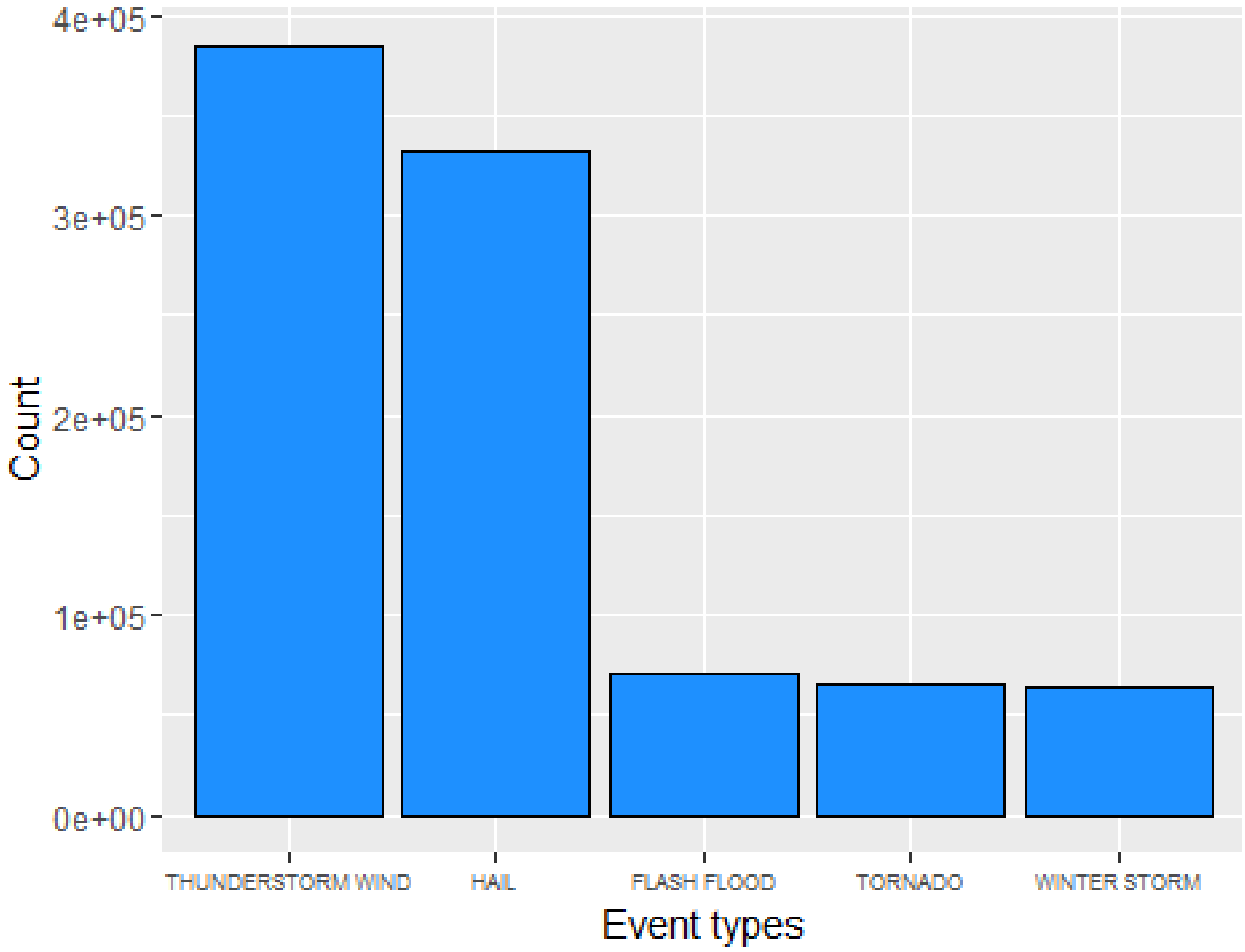

It may be illustrative to visualize the distribution of damage estimates for the most frequent event, namely 'thunderstorm wind'.

Figure 7 - Distribution of property damages for 'thunderstorm wind' eve

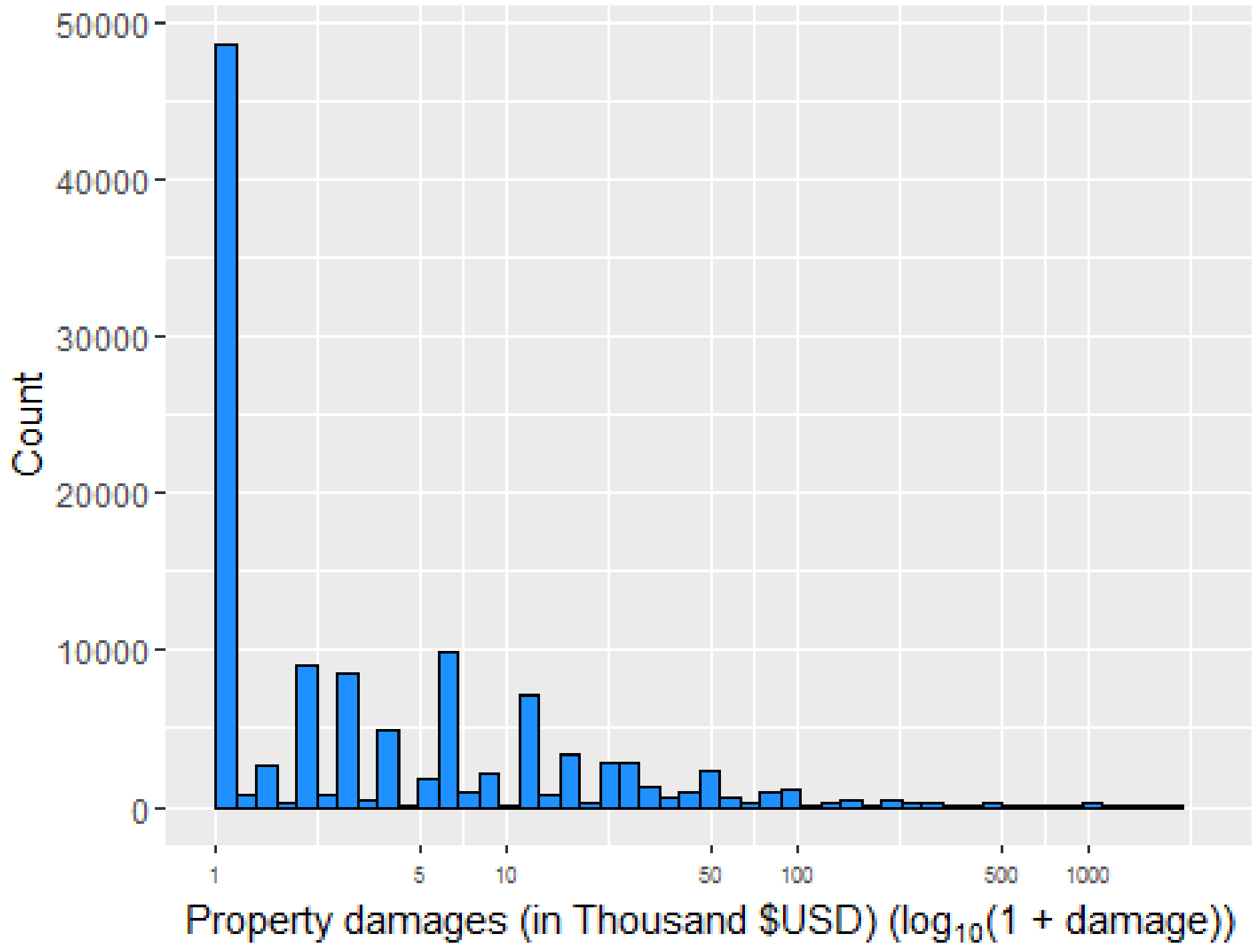

The graph in Figure 7 shows that most damage estimates for 'thunderstorm wind' events are concentrated at $0 value. Inspection shows that 48461 'thunderstorm wind' events have $0 value, what, on top of the high count of 62% of missing values in the *CropDamage* variable in Table 2, suggests a lack of information, as it is difficult to believe that so many events of this kind could result in no damage at all.

As an example, Curran, Holle, & López (2000) concluded that "damage reports appear to be poorly represented" in the *Storm Events Database* as a review in 1989 of 106 entries with damage values of over $500 million showed they to be erroneously coded events, which were then changed to the 'unknown' category. With regard to fatalities, Dixon et al. (2005) concluded that "depending on the database used and the compiling U.S. agency, completely different results can be obtained." As an example, these authors affirm that "there are several instances in Storm Data where traffic-related deaths were classified as directly caused by weather in contrast to the official guidelines."

## Conclusion

NOAA's *Storm Events Database* is undoubtedly an invaluable resource to the general public, to the professional, and to the researcher.

A complete investigation of the damage estimates, injuries or fatalities figures is unfeasible due to the extension of the database.

The few results obtained here, however, are enough to show that the database suffers from incompleteness and inconsistencies and should not be used without taking reservations and appropriate precautions before advancing any inferences from the data.

## Appendix: Code

```r
# Explore NOAA's Storm Database webpage to extract files URL's
NOAAURL <-
"http://www1.ncdc.noaa.gov/pub/data/swdi/stormevents/csvfiles/"
NOAAWPage <- htmlTreeParse(NOAAURL, useInternalNodes = TRUE)
```

```r
# Extract all URL's in NOAA's Storm Database webpage
URLList <- data.frame(fileURL = matrix(unlist(xpathApply(NOAAWPage,
                                              "//a",
                                              function(x) xmlGetAttr(x,
"href"))),
                                        byrow = TRUE),
                      row.names = NULL, stringsAsFactors = FALSE)

# Subset the event details .csv files
URLEDetailsList <- subset(URLList,
                          grepl("StormEvents_details",
                                       URLList$fileURL))

# Auxiliary function to download a file from the website if it has not
been downloaded already
NOAADwnld <- function(file) if (!file.exists(file))
        download.file(paste0(NOAAURL, file), file)

# Download the event details .csv files
dwnldCheck <- rapply(as.list(URLEDetailsList$fileURL), NOAADwnld)

# Check if all files have been downloaded
if(length(dwnldCheck) > 0) message(
        paste(list(URLEDetailsList$fileURL[which(dwnldCheck != 0)]),
              "files were not downloaded!"))

# Read in the datasets from files
# They are in *.bz2* compressed files format but, *read_csv()* from
*readr* package, which is much faster than *read.csv()* (Peng, 2016b, p.
28), can handle them automatically
# The only relevant variables for this study are EPISODE_ID, YEAR,
MONTH_NAME, EVENT_TYPE, DAMAGE_PROPERTY, DAMAGE_CROPS, EPISODE_NARRATIVE
message("Reading data. This may take some time. Please be patient.")

NOAARead <- function(fileName) read_csv(fileName,
                                        col_names = TRUE,
                                        cols_only(EPISODE_ID =
col_character(),
                                                  YEAR = col_integer(),
                                                  MONTH_NAME =
col_character(),
                                                  EVENT_TYPE =
col_character(),
                                                  DAMAGE_PROPERTY =
col_character(),
                                                  DAMAGE_CROPS =
col_character(),
                                                  EPISODE_NARRATIVE =
col_character()))
```

```r
# Join the datasets into one
stormData <- NULL

for(i in 1:nrow(URLEDetailsList)) {
        stormData <- rbind(stormData,
                  NOAARead(URLEDetailsList$fileURL[i]))
}

# Make the variable names more readable, for convenience
stormData <- rename(stormData,
                    EpisodeID = EPISODE_ID,
                    Year = YEAR,
                    Month = MONTH_NAME,
                    EventType = EVENT_TYPE,
                    PropertyDamage = DAMAGE_PROPERTY,
                    CropDamage = DAMAGE_CROPS,
                    Narrative = EPISODE_NARRATIVE)

stormData$EventType <- toupper(stormData$EventType)

# Plot the introduction of non-standard event names into the database
allEventTypes <- toupper(c("Astronomical Low Tide", "Avalanche",
"Blizzard", "Coastal Flood", "Cold/Wind Chill", "Debris Flow",
"Dense Fog", "Dense Smoke", "Drought", "Dust Devil", "Dust Storm",
"Excessive Heat", "Extreme Cold/Wind Chill", "Flash Flood", "Flood",
"Frost/Freeze", "Funnel Cloud", "Freezing Fog", "Hail", "Heat",
"Heavy Rain", "Heavy Snow", "High Surf", "High Wind",
"Hurricane (Typhoon)", "Ice Storm", "Lake-Effect Snow",
"Lakeshore Flood", "Lightning", "Marine Dense Fog", "Marine Hail",
"Marine Heavy Freezing Spray", "Marine High Wind",
"Marine Hurricane/Typhoon", "Marine Lightning",
"Marine Strong Wind", "Marine Thunderstorm Wind",
"Marine Tropical Depression", "Marine Tropical Storm",
"Rip Current", "Seiche", "Sleet", "Sneaker Wave", "Storm Surge/Tide",
"Strong Wind", "Thunderstorm Wind", "Tornado", "Tropical Depression",
"Tropical Storm", "Tsunami", "Volcanic Ash", "Waterspout", "Wildfire",
"Winter Storm", "Winter Weather"))

totEventType <- aggregate(EventType ~ Year,
                          data = stormData,
                          FUN = function(x) x)

totEventType$EventNumber <- rapply(as.list(totEventType$EventType),
                                   length)

testTypes <- function(LofEvTypes) {
        EventTypes <- unlist(LofEvTypes)
        diffEventTypes <- list()
        for(j in 1:length(EventTypes)) {
```

```r
                if(!(EventTypes[j] %in% allEventTypes))
                        diffEventTypes <-
                                append(diffEventTypes, EventTypes[j])
                }
        list(as.vector(unlist(diffEventTypes)))
        }

# Unique event types
uniEventType <- totEventType

for(i in 1:nrow(uniEventType)) {
        uniEventType$EventType[i] <-

list(as.vector(unique(unlist(totEventType$EventType[i]))))
}

uniEventType$EventNumber <- rapply(as.list(uniEventType$EventType),
                                   length)

difuEventType <- uniEventType

for(i in 1:nrow(difuEventType)) {
        difuEventType$EventType[i] <-
                testTypes(difuEventType$EventType[i])
        difuEventType$EventNumber[i] <-
                length(unlist(difuEventType$EventType[i]))
        }

plot1 <- ggplot(data = difuEventType,
       aes(x = Year,
           y = EventNumber))
plot1 <- plot1 + geom_bar(stat = "identity",
                          fill = "dodgerblue",
                          color="black")
plot1 <- plot1 + scale_x_continuous(breaks = c(1950, 1955, 1960,
                                               1965, 1970, 1975,
                                               1980, 1985, 1990,
                                               1995, 2000, 2005,
                                               2010, 2015))
plot1 <- plot1 + scale_y_continuous(breaks = c(0, 2, 4, 6, 8, 10, 12))
plot1 <- plot1 + theme(axis.text.x = element_text(size = 8,
                                                  angle = 45,
                                                  vjust = 0.5,
                                                  hjust=1),
                               plot.title = element_text(size = 10,
                                                         face = "bold"))
plot1 <- plot1 + ylab("Yearly totals")
plot1 <- plot1 + xlab("Year")
plot1 <- plot1 + ggtitle("Figure 1 - Number of non-standard event names
```

```
introduced")
plot1

# Plot the cumulative event types
cumEventType <- uniEventType

for(i in 2:nrow(cumEventType))
        for (j in 1:i-1)
                cumEventType$EventType[i] <-
        list(unique(c(unlist(cumEventType$EventType[i]),
                      unlist(cumEventType$EventType[j]))))

cumEventType$EventNumber <- rapply(as.list(cumEventType$EventType),
                                   length)

plot2 <- ggplot(data = cumEventType,
                aes(x = Year,
                    y = EventNumber))
plot2 <- plot2 + geom_bar(stat = "identity",
                          fill = "dodgerblue",
                          color="black")
plot2 <- plot2 + scale_x_continuous(breaks = c(1950, 1955, 1960,
                                               1965, 1970, 1975,
                                               1980, 1985, 1990,
                                               1995, 2000, 2005,
                                               2010, 2015))
plot2 <- plot2 + scale_y_continuous(breaks = c(0, 20, 40, 48, 55, 60, 80,
100))
plot2 <- plot2 + theme(axis.text.x = element_text(size = 8,
                                                  angle = 45,
                                                  vjust = 0.5,
                                                  hjust=1),
                       plot.title = element_text(size = 10,
                                                 face = "bold"))
plot2 <- plot2 + geom_hline(yintercept = 48)
plot2 <- plot2 + geom_hline(yintercept = 55)
plot2 <- plot2 + ylab("Cumulative totals")
plot2 <- plot2 + xlab("Year")
plot2 <- plot2 + ggtitle("Figure 2 - Number of distinct event names")
plot2

# Tabulate the percentage of non-standard event type names in the dataset
diftEventType <- totEventType

for(i in 1:nrow(diftEventType)) {
        diftEventType$EventType[i] <-
            testTypes(diftEventType$EventType[i])
        diftEventType$EventNumber[i] <-
            length(unlist(diftEventType$EventType[i]))
        diftEventType$EventPercent[i] <-
```

```r
            diftEventType$EventNumber[i] /
            totEventType$EventNumber[i]
        }

diftEventType <- select(diftEventType,
                        Year,
                        EventPercent)

plot3 <- ggplot(data = diftEventType,
                aes(x = Year,
                    y = EventPercent))
plot3 <- plot3 + geom_bar(stat = "identity",
                          fill = "dodgerblue",
                          color="black")
plot3 <- plot3 + scale_x_continuous(breaks = c(1950, 1955, 1960,
                                               1965, 1970, 1975,
                                               1980, 1985, 1990,
                                               1995, 2000, 2005,
                                               2010, 2015))
plot3 <- plot3 + scale_y_continuous(labels = percent_format())
plot3 <- plot3 + theme(axis.text.x = element_text(size = 8,
                                                  angle = 45,
                                                  vjust = 0.5,
                                                  hjust=1),
                       plot.title = element_text(size = 10,
                                                 face = "bold"))
plot3 <- plot3 + ylab("Percentage")
plot3 <- plot3 + xlab("Year")
plot3 <- plot3 + ggtitle("Figure 3 - Percentages of non-standard event
type names in the database.")
plot3

# Recalculate the variables PropertyDamage and CropDamage taking these
magnitude values into account
multipliers <- matrix(data=c(c("0", 1),
                             c("1", 10),
                             c("2", 1e+02),
                             c("3", 1e+03),
                             c("4", 1e+04),
                             c("5", 1e+05),
                             c("6", 1e+06),
                             c("7", 1e+07),
                             c("8", 1e+08),
                             c("H", 1e+02),
                             c("h", 1e+02),
                             c("K", 1e+03),
                             c("k", 1e+03),
                             c("M", 1e+06),
                             c("m", 1e+06),
                             c("B", 1e+09),
```

```
                                       c("b", 1e+09)),
                                nrow=17, ncol=2, byrow = TRUE)

stormData <- mutate(stormData,
                    PropertyDamage =
                            ifelse(is.na(PropertyDamage), NA,
                                   as.numeric(substr(PropertyDamage,
                                                     1,

nchar(PropertyDamage) - 1)) *
                                   as.numeric(multipliers[match(
                                           substr(PropertyDamage,
                                          nchar(PropertyDamage),
                                          nchar(PropertyDamage)),
                                      multipliers[, 1]), 2])),
                    CropDamage =
                            ifelse(is.na(CropDamage), NA,
                                   as.numeric(substr(CropDamage,
                                                      1,
                                        nchar(CropDamage) - 1)) *
                                   as.numeric(multipliers[match(
                                           substr(CropDamage,
                                          nchar(CropDamage),
                                          nchar(CropDamage)),
                                      multipliers[, 1]), 2])))

# Tabulate missing values (NA) in the database
missingValues <- colSums(is.na(stormData))/nrow(stormData)
missingValues <- data.frame(Variable = names(missingValues),
                            MissingValues = missingValues,
                            row.names = NULL)

missingValues$Variable <- factor(missingValues$Variable,
                                 levels =
missingValues$Variable[order(missingValues$MissingValues,

decreasing = TRUE)])
missingValues <- missingValues[order(missingValues$MissingValues,
                                     decreasing = TRUE),]

plot4 <- ggplot(data = missingValues,
                aes(x = Variable,
                    y =  MissingValues))
plot4 <- plot4 + geom_bar(stat = "identity",
                          fill = "dodgerblue",
                          color="black")
plot4 <- plot4 + scale_y_continuous(labels = percent_format())
plot4 <- plot4 + theme(axis.text.x = element_text(size = 8,
                                                   angle = 45,
                                                   hjust=1),
```

```r
                        plot.title = element_text(size = 10,
                                                  face = "bold"))
plot4 <- plot4 + ylab("Missing values")
plot4 <- plot4 + xlab("Variables")
plot4 <- plot4 + ggtitle("Figure 4 - Percentages of missing values in the
variables.")
plot4

# Summarize the database

damages <- na.exclude(rbind(
    data.frame(Damage = stormData$PropertyDamage,
               Type = "Property"),
    data.frame(Damage = stormData$CropDamage,
               Type = "Crop")))

plot5 <- ggplot(data = damages,
                aes(factor(Type),
                    1 + Damage))
plot5 <- plot5 + geom_violin(adjust = 3,
                             fill = "dodgerblue",
                             na.rm = TRUE)
plot5 <- plot5 + geom_boxplot(outlier.colour = "red",
                              outlier.shape = 1,
                              width=0.02,
                              na.rm = TRUE)
plot5 <- plot5 + scale_y_log10(breaks = c(1e-01, 1e+01, 1e+03,
                                          1e+05, 1e+07, 1e+09,
                                          1e+11, 1e+13))
plot5 <- plot5 + theme(axis.text.x = element_text(size = 8,
                                                  vjust = 0.5),
                        plot.title = element_text(size = 10,
                                                  face = "bold"))
plot5 <- plot5 + xlab("Damage type")
plot5 <- plot5 + ylab("Damage value ($USD)")
plot5 <- plot5 + ggtitle("Figure 5 - Distribution of damage values")
plot5

# Introduce TotDamage variable accounting for the total damage.
stormData <- mutate(stormData,
                    TotDamage = ifelse(is.na(CropDamage), NA,
                                ifelse(is.na(PropertyDamage), NA,
                                       PropertyDamage + CropDamage)))

topDamEpisodes <- aggregate(x = stormData["TotDamage"],
                            by = list(EpisodeID = stormData$EpisodeID,
                                      Year = stormData$Year,
                                      Month = stormData$Month),
                            FUN = sum)
```

```
topDamEpisodes$TotDamage <- topDamEpisodes$TotDamage/1e+09

topDamEpisodes <- topDamEpisodes[order(topDamEpisodes$TotDamage,
                                       decreasing = TRUE),]

topDamEpisodes <- head(topDamEpisodes, 10)

row.names(topDamEpisodes) <- NULL

pandoc.table(topDamEpisodes,
             caption = "The top 10 costliest events (in Billion $USD)",
             digits = 1,
             round = 1,
             keep.line.breaks = TRUE,
             justify = c("right", "right", "center", "right"))

topEvents <- aggregate(x = stormData["EventType"],
                       by = list(stormData$EventType),
                       FUN = NROW)

colnames(topEvents) <- c("EventType", "Count")

topEvents$Count <- as.integer(topEvents$Count)

topEvents$EventType <- factor(topEvents$EventType,
                              levels =
topEvents$EventType[order(topEvents$Count,

decreasing = TRUE)])

topEvents <- topEvents[order(topEvents$Count, decreasing = TRUE), ]

plot6 <- ggplot(data = head(topEvents, 5),
                aes(x = EventType,
                    y =  Count))
plot6 <- plot6 + geom_bar(stat = "identity",
                          fill = "dodgerblue",
                          color="black")
plot6 <- plot6 + theme(axis.text.x = element_text(size = 6,
                                                  vjust = 0.5),
                       plot.title = element_text(size = 10,
                                                 face = "bold"))
plot6 <- plot6 + ylab("Count")
plot6 <- plot6 + xlab("Event types")
plot6 <- plot6 + ggtitle("Figure 6 - The 5 most frequent event types in
the database.")
plot6

topEvent <- na.omit(select(filter(stormData[order(stormData$TotDamage,
                                                  decreasing = TRUE),],
```

```
                                   EventType == topEvents$EventType[1]),
                            EventType, TotDamage))

plot7 <- ggplot(data = topEvent, aes(1 + topEvent$TotDamage/1e+03))
plot7 <- plot7 + geom_histogram(color="black",
                                   boundary = 0,
                                   fill = "dodgerblue",
                                   bins = 50,
                                   na.rm = TRUE)
plot7 <- plot7 + scale_x_log10(limits = c(1, 2500),
                                breaks=c(1, 5, 10, 50, 100,
                                         500, 1000))
plot7 <- plot7 + theme(axis.text.x = element_text(size = 6,
                                                   vjust = 0.5),
                        plot.title = element_text(size = 10,
                                                  face = "bold"))

plot7 <- plot7 + xlab(expression("Property damages (in Thousand $USD)
(log"[10]*"(1 + damage))"))
plot7 <- plot7 + ylab("Count")
plot7 <- plot7 + ggtitle(paste0("Figure 7 - Distribution of property
damages for '",
                                    tolower(topEvent$EventType[1]), "'
events."))
plot7
```

## Affiliation

Renato P. dos Santos

ULBRA - Lutheran University of Brazil

PPGECIM - Doctoral Program in Science and Mathematics Education

Av. Farroupilha, 8001 - 92425-900 Canoas/RS - Brazil

E-mail: renatopsantos@ulbra.edu.br

URL: http://www.linkedin.com/in/RenatoPdosSantos